\documentclass[11pt,oneside]{article}
\usepackage{a4wide}
\usepackage{mathrsfs}
\usepackage{latexsym,bm}
\usepackage{graphicx}
\usepackage{indentfirst}
\usepackage{slashed}
\usepackage{amsmath}
\usepackage{amssymb}
\usepackage{color}
\usepackage{hyperref}
\usepackage{epsfig}
\usepackage[titletoc]{appendix}
\usepackage{multirow}%
\usepackage{rotating}
\usepackage{cite}
\usepackage{ulem} %% \sout
\usepackage[top=2.9cm,bottom=2.5cm,left=2.8cm,right=3cm]{geometry}
\usepackage{tabularx}
\newcommand{\be}{\begin{equation}} \newcommand{\ee}{\end{equation}}
\newcommand{\ba}{\left(\begin{array}{c}}
\newcommand{\ea}{\end{array}\right)}
\newcommand{\bea}{\begin{eqnarray}} \newcommand{\eea}{\end{eqnarray}}

\newcommand{\email}[1]{\footnote{ \texttt{#1}}}

\newcommand{\bma}{\left(\begin{matrix}}
\newcommand{\ema}{\end{matrix}\right)}
\newcommand{\bqa}{\begin{eqnarray}}
\newcommand{\eqa}{\end{eqnarray}}
\newcommand{\bqaa}{\begin{eqnarray*}}
\newcommand{\eqaa}{\end{eqnarray*}}

\newcommand{\nn}{\nonumber}

\begin{document}
%\begin{CJK}{GBK}{kai}
\thispagestyle{empty}
\title{%\vspace{-3cm}{\small {\color{red} {NOTE \hfill version 2.0}}}\\[2cm]
\Large \bf Effective-range-expansion study of near threshold heavy-flavor resonances}
\author{\small Rui Gao$^{a}$\email{1416649615@qq.com}, \, Zhi-Hui Guo$^{a}$\email{zhguo@hebtu.edu.cn}, \, Xian-Wei Kang$^{b}$\email{11112018023@bnu.edu.cn}, \, J.~A.~Oller$^{c}$\email{oller@um.es}  \\[0.5em]
{ \small\it ${}^a$  Department of Physics and Hebei Advanced Thin Films Laboratory, } \\
{\small\it Hebei Normal University,  Shijiazhuang 050024, China}\\[0.3em]
{ \small\it ${}^b$ College of Nuclear Science and Technology, Beijing Normal University, Beijing 100875, China }\\[0.3em]
{\small {\it $^c$Departamento de F\'{\i}sica. Universidad de Murcia. E-30071 Murcia. Spain}}
}
\date{}
\maketitle
\begin{abstract}

In this work we study the resonances near the thresholds of the open
heavy-flavor hadrons using the effective-range-expansion method. The
unitarity, analyticity and compositeness coefficient are also taken
into account in our theoretical formalism. We consider the
$Z_c(3900)$, $X(4020)$, $\chi_{c1}(4140)$, $\psi(4260)$ and
$\psi(4660)$. The scattering lengths and effective ranges from the
relevant elastic $S$-wave scattering amplitudes are determined.
Tentative discussions on the inner structures of the aforementioned
resonances are given.

\end{abstract}
% {\small PACS numbers: 12.39.Fe, 13.75.Lb, 14.40.Lb\\
% Key words: Chiral Lagrangian, charmed mesons, meson-meson interaction}

%\\
%\end{@twocolumnfalse}
%]

%\newpage

%\tableofcontents
%\listoftables
%\listoffigures
%\newpage

\section{Introduction}

Since the discovery of the $X(3872)$~\cite{Choi:2003ue} the study of
the exotic hidden heavy-flavor hadrons has become one of the most
important and active research topics in particle physics. Up to now
more than twenty of the so-called  $XYZ$ hadrons are observed by
experiments~\cite{Tanabashi:2018oca} and we refer to
Refs.~\cite{Chen:2016qju,Guo:2017jvc} for recent comprehensive
reviews on this subject. One of the most important features of the
newly observed hadrons is that they are usually close to the nearby
thresholds of the open heavy-flavor states. As a result typically
one needs to properly take into account the threshold effects when
studying those possible exotic states. The effective range expansion
(ERE), which is based on the three-momentum expansion near the
threshold, provides a useful tool to address the dynamics in the
energy region  around the relevant threshold in
question~\cite{021015.1.bethe,021015.2.preston}.

By combining the ERE, unitarity, analyticity and the compositeness coefficients developed in Refs.~\cite{Guo:2015daa,Oller:2017alp},
we have successfully analyzed several non-ordinary hadronic states that lie close to the thresholds of the underlying two-particle states,
including the $\Lambda_c(2595)$~\cite{Guo:2016wpy}, $Z_b(10610)$ and $Z_b(10650)$~\cite{Kang:2016ezb}, 
and the newly observed pentaquark candidates $P_c(4312), P_c(4440)$ and $P_c(4457)$~\cite{Guo:2019kdc}. 
The essential idea of the formalism is 
that we construct the elastic unitarized partial-wave amplitude using the ERE as the kernel, which includes the scattering length $a$ and
effective range $r$ as free parameters.
The latter are determined by reproducing the values of the mass and width of the observed resonance. %O-1%
 We always obtain real values consistently with the assumption of only one-channel scattering. %O-1%
 Then the residue of the resonance pole, which corresponds to the coupling strength of the resonance to the interacting two-particle state,
 can be obtained as well. With all of these ingredients, we can apply the compositeness formalism to infer the probability to find the
  two-particle state inside the resonance. In this work we first briefly recapitulate the essentials of the theoretical formalism.
  Then we tentatively generalize this formalism to other newly observed hadronic states,
  including the $Z_c(3900)$ near the $D\bar{D}^*+c.c.$ (denoted shortly as $D\bar{D}^*$ in the following) threshold,
  the $X(4020)$ near the $D^*\bar{D}^*$ threshold, the $\chi_{c1}(4140)$ near the $D_s^{\pm} D_s^{*\mp}$
 (denoted as $D_s\bar{D}_s^*$ in the following) threshold, the $\psi(4260)$ near the $D_1\bar{D}+c.c.$
 (denoted shortly as $D_1\bar{D}$ in the following) threshold, and the $\psi(4660)$ near the $\Lambda_c \bar{\Lambda}_c$ threshold~\cite{Tanabashi:2018oca}.
  Historically, there is also a state named $X(4630)$ that we identify
with  the resonance $\psi(4660)$ as in the PDG \cite{Tanabashi:2018oca} and
Refs.~\cite{Dai,Polosa,Guo}.

\section{Effective range expansion and compositeness coefficients of resonances}

The basic staring point of our theoretical formalism is the ERE up to the next-to-leading order
\begin{eqnarray}\label{eq.ti}
t(E)=\frac{1}{-\frac{1}{a}+\frac{1}{2}r\,k^2-i\,k}\,,
\end{eqnarray}
with $a$ the scattering length, $r$ the effective range and $k$ the three-momentum in the center of mass (CM) frame.
At a given CM energy $E$ around the threshold, one can write the nonrelativistic three-momentum as
\begin{equation}\label{eq.ek}
k=\sqrt{2\mu_m(E-m_{\rm th})}\,,
\end{equation}
where the reduced mass of the system with masses $m_1$ and $m_2$ is $\mu_m=m_1 m_2/(m_1+m_2)$ and the threshold is given by $m_{\rm th}=m_1+m_2$.

We mention that a more general expression to write the scattering amplitude near threshold is to include the so-called Castillejo-Dalitz-Dyson (CDD) poles. The standard ERE in Eq.~\eqref{eq.ti} can be obtained by expanding the full expression with CDD poles~\cite{Guo:2016wpy,Kang:2016ezb}.
However when the CDD pole happens to be near the threshold, the expansion in terms of $k^2$ will be invalid, or at least quite limited to a tiny region. One of the prominent features in this situation is the huge effective range $r$, which usually becomes much larger than its standard value around 1~fm. If this is the case, one has to work explicitly with the CDD pole in the full expression, as done in Refs.~\cite{Guo:2016wpy,Kang:2016ezb,oller.181211.1}.

The partial-wave amplitude given in Eq.~\eqref{eq.ti} corresponds to the physical one in the first Riemann sheet (RS). Its expression in the second RS is given by
\begin{eqnarray}\label{eq.tii}
t_{\rm II}(E)=\frac{1}{-\frac{1}{a}+\frac{1}{2}r\,k^2+i\,k}\,.
\end{eqnarray}
The resonance poles only appear in the second RS. Comparing with Eqs.~\eqref{eq.ti} and \eqref{eq.tii}, there is a change of sign for the $k$ term. Notice that in the conventions of Eqs.~\eqref{eq.ti} and \eqref{eq.tii} the %imaginary part of 
 three-momentum  $k$ should be evaluated with ${\rm Im}k>0$.

Let us now consider a resonance $R$ whose pole position %O-1%
in the unphysical RS is located at $E=E_R$, with $E_R=M_R-i\Gamma_R/2$.
For a conventional narrow-width resonance, one can identify %$M_R$ as the mass and $\Gamma_R$ as the width.
 $M_R$ and $\Gamma_R$ as its  mass and width, respectively. 
The corresponding three-momentum at the resonance pole is then given by
\begin{eqnarray}
\label{181211.1}
k_R=\sqrt{2\mu_m(E_R-m_{\rm th})}\,.
\end{eqnarray}
For later convenience, we further define
\begin{eqnarray}
k_R=k_r+ik_i\,, \quad k_i>0 \,.
\end{eqnarray}
One has to be careful when evaluating $k_R$ in terms of the threshold $m_{\rm th}$ and the resonance parameters $M_R$ and $\Gamma_R$, specially distinguishing the sign of $M_R-m_{\rm th}$. Detailed discussions can be found in Ref.~\cite{Kang:2016ezb}.

The resonance pole corresponds to the zero of the denominator of $t_{\rm II}(E)$, i.e. at the pole position one has
\begin{eqnarray}
-\frac{1}{a}+\frac{1}{2}r\,k_R^2+i\,k_R=0\,.
\end{eqnarray}
By a straightforward algebraic manipulation, one can solve $a$ and $r$ in terms of $k_r$ and $k_i$
\begin{align}\label{eq.a}
a=&-\frac{2k_i}{|k_R|^2}~,\\
r=&-\frac{1}{k_i}~.\label{eq.r}
\end{align}

Substituting Eqs.~\eqref{eq.a} and \eqref{eq.r} into Eq.~\eqref{eq.tii}, one can write the partial-wave amplitude around the resonance pole $E_R$ as
\begin{eqnarray}\label{eq.tiikr}
t_{II}(k)=\frac{-k_i/k_r}{k-k_R}+\ldots \,,
\end{eqnarray}
where the ellipses stand for the neglected terms when expanding the denominator of Eq.~\eqref{eq.tii} in terms of $k-k_R$. From Eq.~\eqref{eq.tiikr} we can identify the residue at the pole in the variable $k$, which turns out to be
\begin{eqnarray}\label{eq.gammak}
\gamma_k^2=-\frac{k_i}{k_r}>0\,,
\end{eqnarray}
since $k_r<0$. %O-1$
Alternatively one can also expand the partial-wave amplitude around the pole as
\begin{eqnarray}
 t_{II}(E)= -\frac{\gamma^2}{s-E_R^2} +\ldots\,.
\end{eqnarray}
The residue $\gamma^2$ is related to $\gamma_k^2$ as
\begin{eqnarray}
\gamma_k^2=-\gamma^2 \left.\frac{dk}{ds}\right|_{k_R}=-\frac{\mu_m \gamma^2}{2E_R k_R}\,.
\end{eqnarray}

In Ref.~\cite{Guo:2015daa}, a probabilistic interpretation for the compositeness $X$ of the
two-particle state inside the resonance is derived. The value of $X$ can be calculated once
the resonance pole position and the corresponding residues are provided. %O-1%
Around the two-particle threshold, the probability $X$ reduces to ~\cite{Kang:2016ezb}
\begin{eqnarray}\label{eq.x}
 X = |\gamma_k|^2\,.
\end{eqnarray}
However, we point out that the probabilistic interpretation of $X$ is restricted to
resonances with $M_R>m_{\rm th}$~\cite{Guo:2015daa}.  %O-1% 
In Refs.~\cite{Baru:2003qq,Hanhart:2011jz,Hyodo:2011qc,Aceti:2012dd,Sekihara:2014kya,hyodo.190429.1,hyodo.190429.2,sekihara.190429.1,aceti.190430.1}, other approaches to generalize the Weinberg's compositeness of bound states~\cite{Weinberg:1962hj} to the resonances are discussed.
We compare below our results with some of them  in Sec.~\ref{sec.190430.1}.

In the next section, we proceed to study several near-threshold resonances within the present ERE approach.
Let us notice that  if  this type of ERE study is applied to a near-threshold resonance 
which is composite of the nearby channel (the so-called elastic one)
then $X\simeq 1$. From here it also follows that if we apply this type of ERE study to a near-threshold resonance and
it results that $X\ll  1$,   then one can conclude for sure that this resonance is not a composite one 
of the elastic channel.
On the other hand, if it results that $X\simeq 1$ then the interpretation of this
resonance as a composite one of the elastic channel is favored.~\footnote{In a similar sense that a cloudy
  sky favors rainfall.}

%%%%%%%%%%%%%%%%%%%%%%%%%%%%%%%%%%%%%%%%%%%%%%%%%%%%%%%%%%%%%%%%%%%%%%%%%%%%%%%%%%%%%%%%%%%%%%%%%%%%%%%%%
\section{Phenomenological discussions}
\label{sec.190430.1}

Before entering the detailed discussions,
we stress that the theoretical approach developed is based on the elastic $S$-wave
two-body scattering. %O-1% 
Strictly speaking, the present formalism applies %strictly 
to the %stable two-hadron elastic scattering.
scattering of two stable hadrons. 
A rigorous approach to handle %
the presence of unstable hadrons in the scattering process
is to perform the study within three- or 
even %
few-body scattering \cite{baru.190330.1}, which is clearly beyond the scope of the present work. 
Another indirect way to understand the role of the width in the unstable hadrons is to use a
complex mass ($m_i\to m_i-i\Gamma_i/2$) in the expression
for the three-momentum, Eq.~\eqref{181211.1}, which then reads 
\begin{align}
\label{190330.1}
k_R=\sqrt{2\mu_m\left(E_R-m_{\rm th}+i\frac{\Gamma}{2}\right)}\,.  
\end{align} 
In this way, one can take into account the self-energy effects of the decaying channels.
This is applicable, as compared with a full three-body study,  if $\Gamma$ is much smaller than the difference between
the mass of the resonance and the threshold of the decay channel, as indicated by the results from
Ref.~\cite{baru.190330.1} concerning the $D\bar{D}^*$, $\bar{D}D^*$ and $D\bar{D}\pi$ scattering
and the $X(3872)$ resonance. This is
the case  for the $D_1$ hadron, and the numerical results obtained with a complex mass  are also indicated
below. 
 Another interesting point is that unstable hadrons could introduce additional crossed-channel contributions,
 comparing with stable ones, see e.g. \cite{baru.190330.1}.

The crossed-channel effects, such as the light-flavor hadron exchanges, are  neglected in Eq.~\eqref{eq.ti}, which is strictly valid in the near-threshold region before any other branch-point singularity, associated with the onset of crossed channels, sets in.
Nonetheless, the theoretical formalism presented here can be  used also to study the systems in which the crossed-channel
dynamics can be treated perturbatively. %O-1% 
In Refs.~\cite{Guo:2016wpy,Kang:2016ezb}, the resonances $\Lambda_c(2595)$, $Z_b(10610)$ and $Z_b(10650)$
have been successfully addressed within this formalism.
In the following we tentatively generalize the discussions to the $Z_c(3900)$, $X(4020)$, $\chi_{c1}(4140)$, $\psi(4260)$ and $\psi(4660)$, which may be composed by some specific $S$-wave open-charm two-body states and lie close to their thresholds.

The masses and widths of the $Z_c(3900)$, $X(4020)$, $\chi_{c1}(4140)$, $\psi(4260)$ and $\psi(4660)$, and the thresholds of the nearby two-body states are collected in Table~\ref{tab.arx}. The spin and parity of the $Z_c(3900)$, $\chi_{c1}(4140)$, $\psi(4260)$ and $\psi(4660)$ given
in the PDG~\cite{Tanabashi:2018oca} are compatible with the $S$-wave elastic scattering of $D\bar{D}^*$, $D_s\bar{D}_s^*$, $D_1\bar{D}$ and $\Lambda_c\bar{\Lambda}_c$, respectively. For the $X(4020)$, its spin and parity are not confirmed by experiments yet. By assuming the $S$-wave molecular nature of $D^*\bar{D}^*$, a possible quantum number $J^{PC}$ of the $X(4020)$ would be $1^{+-}$.

%%%%%%%%%%%%%%%%%%%%%%%%%%%%%%%%%%%%%%%%%%%%%%%%%%%%%%%%%%%%%%%%%%%%%%%%%%%%%%%%%%%%%%%%%%%%%%%%%%%%%%%%%%%%%%%%%%
%%%%%%%%%%%%%%%%%%%%%%%%%%%%%%%%%%%%%%%%%%%%%%%%%%%%%%%%%%%%%%%%%%%%%%%%%%%%%%%%%%%%%%%%%%%%%%%%%%%%%%%%%%%%%%%%%%
\begin{table}[htbp]
\centering
\caption{\label{tab.arx} In the second and third columns, the masses and widths of the $Z_c(3900)$, $X(4020)$,  $\psi(4260)$ and $\psi(4660)$ from the PDG are given. For the $\chi_{c1}(4140)$, we have distinguished three different determinations: LHCb, the average without LHCb (\sout{LHCb}) and the PDG average. To make a conservative error estimate, the largest error bars are taken for the asymmetric ones in the values from the LHCb and PDG. We assume that the $S$-wave two-particle states shown in the fourth column are responsible for the resonance poles. The corresponding thresholds are also explicitly given. The elastic scattering lengths, effective ranges and the compositeness coefficients are provided in the last three columns. Since the mass of the $\psi(4260)$ is below the $D_1\bar{D}$ threshold, the probabilistic interpretation of $X$ does not hold in this situation~\cite{Guo:2015daa}.} 
\begin{scriptsize}
%\begin{ruledtabular}
\begin{tabular}{ c c c c c c c}
\hline\hline
 Resonance & Mass   & Width & Threshold  & $a$     & $r$    & $X$  \\
           & (MeV)  & (MeV) &  (MeV)    & (fm)    &  (fm)      &
\\ \hline
$Z_c(3900)$ & $3886.6\pm 2.4$ & $28.2\pm 2.6$ & $D\bar{D}^*$~(3875.8) & $-0.94 \pm 0.12$ & $-2.40 \pm0.21$ & $0.49 \pm 0.06$
\\ \hline
$X(4020)$ & $4024.1\pm 1.9$ & $13\pm 5$  & $D^*\bar{D}^*$~(4017.1) & $-1.04 \pm0.26$ & $-3.89 \pm 1.42$ & $0.39 \pm 0.12$
\\ \hline
$\psi(4260)$ & $4230\pm 8$ & $55\pm 19$ & $D_1\bar{D}$~(4289.2) & $-1.04 \pm 0.06$ & $-0.54 \pm 0.03$ & $---$
\\ \hline
$\psi(4660)$ & $4643\pm 9$ & $72\pm 11$ & $\Lambda_c\bar{\Lambda}_c$~(4572.9) & $-0.22 \pm 0.04$ & $-1.98 \pm 0.28$ & $0.24 \pm 0.04$
\\ \hline
$\chi_{c1}(4140)$ (LHCb) & $4146.5\pm 6.4$ & $83\pm 30$ & $D_s\bar{D}_s^*$~(4080.5) & $-0.27 \pm 0.06$ & $-1.79 \pm 0.61$ & $0.29 \pm0.08$
\\
$\chi_{c1}(4140)$ (\sout{LHCb}) & $4147.1\pm 2.4$ & $15.7\pm 6.3$ & $D_s\bar{D}_s^*$~(4080.5) & $-0.06 \pm 0.02$ & $-9.10 \pm 3.86$ & $0.06 \pm 0.02$
\\
$\chi_{c1}(4140)$ (PDG) & $4146.8\pm 2.4$ & $22\pm 8$ & $D_s\bar{D}_s^*$~(4080.5) & $-0.09 \pm0.03$ & $-6.49 \pm 2.40$ & $0.08 \pm 0.03$
\\
\hline\hline
\end{tabular}
%\end{ruledtabular}
\end{scriptsize}
\end{table}
%%%%%%%%%%%%%%%%%%%%%%%%%%%%%%%%%%%%%%%%%%%%%%%%%%%%%%%%%%%%%%%%%%%%%%%%%%%%%%%%%%%%%%%%%%%%%%%%%%%%%%%%%%%%%%%%%%
%%%%%%%%%%%%%%%%%%%%%%%%%%%%%%%%%%%%%%%%%%%%%%%%%%%%%%%%%%%%%%%%%%%%%%%%%%%%%%%%%%%%%%%%%%%%%%%%%%%%%%%%%%%%%%%%%%

For the $Z_c(3900)$, we see that standard values of $a$ and $r$, corresponding
to the typical scale of the long-range force of strong interactions,
result from the $D\bar{D}^*$ scattering, according to the results in Table~\ref{tab.arx}.
The component of $D\bar{D}^*$ inside the $Z_c(3900)$ is as important as the other degrees of freedom (d.o.f),
according to the compositeness $X\sim 0.5$. This finding qualitatively agrees with the conclusion from the pole-counting-rule study~\cite{Gong:2016hlt}. 
We have used the error bars of the masses and widths of the resonances to estimate the uncertainties of the $a$, $r$ and $X$, and neglected the tiny error bars of the thresholds.
For the $X(4020)$, a somewhat large value for $|r|$ is obtained.
Both the $D^*\bar{D}^*$ and other % hadronic
d.o.f play important roles in the formation of $X(4020)$.

It is proposed in Refs.~\cite{Wang:2013cya,Wang:2013kra} that the $\psi(4260)$ could be a possible $S$-wave $D_1\bar{D}$ resonance. If one assumes that the $S$-wave $D_1\bar{D}$ is the only source which is responsible for the resonance pole,
the resulting scattering length and effective range can be found in Table~\ref{tab.arx}.
Standard values of $a$ and $r$ around 1~fm are obtained.
However,  because the resonance pole of $\psi(4260)$ is below the $D_1\bar{D}$ threshold,
we can not use the recipe in Eq.~\eqref{eq.x} for the probabilistic interpretation~\cite{Guo:2015daa}.
Nonetheless, since the presence of a close to threshold CDD pole is characterized by having a small scattering length and
a big effective range in magnitudes, the natural values for $a$ and $r$ given in Table~\ref{tab.arx} favor the interpretation
offered in Refs.~\cite{Wang:2013cya,Wang:2013kra}. 
As shown in Ref.~\cite{Kang:2016ezb} when the position of the CDD pole tends to threshold,
$M_{\rm CDD}\to m_{\rm th}$, the resulting $a$ and $r$ tend to
\begin{align}
  \label{181211.2}
  a&\to \frac{M_{\rm CDD}-m_{\rm th}}{\lambda}~,\\
  r&\to -\frac{\lambda}{\mu_m(M_{\rm CDD}-m_{\rm th})^2}~,\nn
\end{align}
where $\lambda$ is the residue of the CDD pole. The bigger this
residue is, the sooner this behavior sets in. When the finite width of the $D_1$ is included via Eq.~\eqref{190330.1}, the scattering length $a$ and effective range $r$ %2will be
are shifted to $-1.10\pm 0.07$~fm and $-0.55\pm 0.04$~fm, respectively,
which are compatible with the results shown in Table~\ref{tab.arx} within uncertainties.
Therefore the effects from the finite width of the $D_1$ are indeed %2mild.
small. We also note that in Ref.~\cite{Dai2}, the large coupling of $\psi(4260)$ to $\omega\chi_{c0}$ is pointed out. Employing this coupling Ref.~\cite{Guo:2015daa} obtained
that  $X_{\omega\chi_{c0}} \sim 0.2$.

Due to the closeness of the $\psi(4660)$ to the $\Lambda_c\bar{\Lambda}_c$ threshold
and the compatibility of its quantum numbers with the $S$-wave $\Lambda_c\bar{\Lambda}_c$,
we also explore the possibility that the elastic $S$-wave $\Lambda_c\bar{\Lambda}_c$ scattering is responsible for the $\psi(4660)$ pole. %O-1
However, the small value of compositeness coefficient $X$ in Table~\ref{tab.arx} indicates that the $\Lambda_c\bar{\Lambda}_c$
component plays a minor role and other %hadronic
 d.o.f plays a more important role inside the $\psi(4660)$. %O-1

The quantum numbers of the $S$-wave $D_s\bar{D}_s^*$ scattering is compatible with the preferred $J^{PC}=1^{++}$ of
the $\chi_{c1}(4140)$~\cite{Tanabashi:2018oca}.
Notice that the new determinations of the masses and widths from LHCb~\cite{Aaij:2016iza,Aaij:2016nsc} are obviously
larger than the previous measurements. We explicitly give different values for the masses and widths of
$\chi_{c1}(4140)$ in Table~\ref{tab.arx}. In all cases, it seems that the $D_s\bar{D}_s^*$ component
plays a minor role inside $\chi_{c1}(4140)$.

We have employed the approach of Refs.~\cite{Guo:2015daa,Kang:2016ezb} which
conclude a probabilistic interpretation of $|X|$ for $M_R>m_{{\rm th}}$.
As explained in more detail in Refs.~\cite{Guo:2015daa,Oller:2017alp},
this result is based on a phase redefinition
of the resonance couplings (whose phases are affected by the non-resonant terms around the
pole \cite{weinberg.190429.1}), and on a direct observation of the resonance-peak signal
stemming from its Laurent series for physical values of the energy.
In contrast, Refs.~\cite{hyodo.190429.1,hyodo.190429.2,sekihara.190429.1}
propose some ad-hoc  manipulations on the complex numbers $X$ and
$Z=1-X$ so as to end with positive definite values between 0 and 1.
We compare our results with theirs below. 

The  Refs.~\cite{hyodo.190429.1,hyodo.190429.2} construct  from $|X|$ and $|Z|$ other numbers
$\widetilde{X}$, $\widetilde{Z}$ and $U$, which are defined as
\begin{align}
  \label{190429.2}
%  \widetilde{X}&=\frac{|X|+1-|Z|}{2}~,\\
%  \widetilde{Z}&=\frac{|Z|+1-|X|}{2}~,\nn\\
%  U&=|X|+|Z|-1~.\nn
  \widetilde{X}&=\frac{1}{2}+\frac{|X|-|1-X|}{2}~,\\
  \widetilde{Z}&=\frac{1}{2}+\frac{|1-X|-|X|}{2}~,\nn\\
  U&=|X|+|1-X|-1~.\nn
\end{align}
By construction they fulfill that
$\widetilde{X}+\widetilde{Z}=1$ and $0\leq \widetilde{X},\,\widetilde{Z}\leq 1$.
The parameter $U$ measures the degree of cancellation between the complex numbers $X$ and $Z$, whose
sum is 1.
By geometrical reasoning, see the Fig.~1 of Ref.~\cite{hyodo.190429.2},
$\pm U/2$ is interpreted as the uncertainty in the probabilistic interpretation of $\widetilde{X}$.

In turn, Ref.~\cite{sekihara.190429.1} introduces 
other numbers $\hat{X}$ and $\hat{Z}$ defined by
\begin{align}
  \label{194029.1}
%  \hat{X}&=\frac{|X|}{1+U}~,\\
%  \hat{Z}&=\frac{|Z|}{1+U}~.\nn
  \hat{X}&=\frac{|X|}{1+U}~,\\
  \hat{Z}&=\frac{|1-X|}{1+U}~.\nn
\end{align}
These  numbers also fulfill by construction that $0\leq \hat{X},\,\hat{Z}\leq 1$
and $\hat{X}+\hat{Z}=1$. 
 This reference claims that $\hat{X}$ and $\hat{Z}$ have a probabilistic interpretation
 in connection with the weight of the different continuum states in a resonance if $U\ll 1$.
 It is also noticed that the difference between $\widetilde{X}$ and $\hat{X}$ tends to zero linearly
 in $U$ for $U\to 0$. 

\begin{table}
\centering
\caption{\label{table.190429.1} Set of numbers $|X|$, $|\widetilde{X}|$ \cite{hyodo.190429.2},  $|\hat{X}|$ \cite{sekihara.190429.1} and 
$U$ \cite{hyodo.190429.1,hyodo.190429.2,sekihara.190429.1}  
for the resonances considered in Table~\ref{tab.arx}.  
No criteria for the probabilistic interpretation of the compositeness is 
met in the case of the $\psi(4260)$ resonance.}
\begin{scriptsize}
%\begin{ruledtabular}
\begin{tabular}{ c c c c c c}
\hline\hline
                               & Asymptotic       &       &                 & & \\
Resonance                      & State            & $|X|$ & $\widetilde{X}$ & $\hat{X}$ &  $U$ \\
\hline
$Z_c(3900)$                    & $D\bar{D}^*$     &  $0.49 \pm 0.06$ & $0.18\pm 0.02$  & $0.30\pm 0.02$  & $0.60\pm 0.10$   \\\hline
$X(4020)$                      & $D^*\bar{D}^*$   & $0.39 \pm 0.12$  & $0.16 \pm 0.04$ & $0.27 \pm 0.06$ & $0.47 \pm 0.19$  \\\hline
$\psi(4260)$                   & $D_1\bar{D}$     & $4.5\pm 1.6$     & $0.45 \pm 0.02$ & $0.49\pm 0.01$     & $8.1\pm 3.2   $  \\ \hline
$\psi(4660)$                   & $\Lambda_c\bar{\Lambda}_c$ & $0.24 \pm 0.04$ & $0.11\pm 0.02$ & $0.19\pm 0.03$ & $0.27\pm 0.05$ \\\hline
$\chi_{c1}(4140)$ (LHCb)        & $D_s\bar{D}_s^*$ & $0.29 \pm 0.08$  & $0.12\pm 0.03$ & $0.22\pm 0.05$ & $0.33\pm 0.12$ \\
$\chi_{c1}(4140)$ (\sout{LHCb}) & $D_s\bar{D}_s^*$ & $0.06 \pm 0.02$ & $0.03\pm 0.01$ & $0.06\pm 0.02$ & $0.06\pm 0.02$ \\
$\chi_{c1}(4140)$ (PDG)         & $D_s\bar{D}_s^*$ & $0.08 \pm 0.03$ & $0.04\pm 0.01$ & $0.08\pm 0.03$ & $0.09\pm 0.03$ \\
\hline
\hline
\end{tabular}
\end{scriptsize}
\end{table}

We give in Table~\ref{table.190429.1}
the values of $|X|$, $\widetilde{X}$, $\hat{X}$ and $U$
for the resonances shown in Table~\ref{tab.arx}. For the last two entries of the $\chi_{c1}(4140)$ one has that $U\ll 1$ and
the same quantitative conclusion on the very small size of the compositeness of $D_s \bar{D}_s^*$ results from all these numbers.
Still small values for $U\lesssim 0.3$ results for the first entry of the $\chi_{c1}(4140)$ and for the $\psi(4660)$, and a
similar conclusion on the relatively small size of the weight of the two-body continuum states results from all the instances.
Notice that it is always the case in these examples that $|X|\simeq \hat{X}$, while $\widetilde{X}$ is different by around a factor of 2. 
This is because \cite{hyodo.190429.2} 
   \label{194029.3}
 \begin{align}
   \bigg| |X|-\widetilde{X} \bigg|=\frac{U}{2}~,
 \end{align}
 as it is clear from Eq.~\eqref{190429.2}, and $U$ is as large as $|X|$ in these cases. Thus,
 the approach that we follow here, based on Refs.~\cite{Guo:2015daa,Kang:2016ezb} 
and summarized above, favors the use of $\hat{X}$ of Ref.~\cite{sekihara.190429.1} over  $\widetilde{X}$ of Ref.~\cite{hyodo.190429.2}. 

For the resonance  $X(4020)$ one has that $U\approx 0.5$. Now, the difference between the central values of
$|X|$ and $\hat{X}$ is larger, although they drive to a similar conclusion on the weight of the continuum state.
The value of $U$ is larger  for the $Z_c(3900)$, with $U/2\approx 0.3$, and the uncertainty $U/2$ in the interpretation of the numbers $\hat{X}$ and $\widetilde{X}$, as argued in Ref.~\cite{hyodo.190429.2},
becomes very important, driving to values that differ between each other by around a factor of 2.
Finally, any of the criteria for the probabilistic interpretation of the compositeness of the $\psi(4260)$ cannot be applied since now $U$ is huge.

Let us also mention that the ERE for scattering up to and including the effective range,
like in our study here, drives necessarily to purely imaginary values for $X$,
since $X=i\gamma_k^2=-i k_i/k_r$. 
This invalidates the interpretation of ${\rm Re}\, X$ as the compositeness for the case of a resonance,
as advocated in Ref.~\cite{aceti.190430.1}, 
because it is always equal to zero in our case, no matter the nature of the resonance under study.

%%%%%%%%%%%%%%%%%%%%%%%%%%%%%%%%%%%%%%%%%%%%%%%%%%%%%%%%%%%%%%%%%%%%%%%%%%%%%%%%%%%%%%%%%
\section{Summary and conclusions}

In this work we have combined the effective range expansion, unitarity, analyticity and the compositeness coefficient to study the resonance dynamics around the threshold energy region. We only focus on the elastic $S$-wave scattering throughout. In our formalism, the scattering length, effective range and the compositeness coefficient can be straightforwardly calculated, if the mass and width of the resonance are provided.

We have applied the theoretical formalism to the $Z_c(3900)$, $X(4020)$, $\chi_{c1}(4140)$, $\psi(4260)$ and $\psi(4660)$. The resonance poles are assumed to be generated by the elastic $S$-wave scattering of $D\bar{D}^*$, $D^*\bar{D}^*$, $D_s\bar{D}_s^*$, $D_1\bar{D}$ and $\Lambda_c\bar{\Lambda}_c$, respectively. According to the values in Table~\ref{tab.arx}, we tentatively conclude that both the $D\bar{D}^*$ and other %hadronic
degrees of freedom are equally important inside the $Z_c(3900)$. The $D^*\bar{D}^*$ component inside the $X(4020)$ is also important.
While for the  $\chi_{c1}(4140)$ and $\psi(4660)$, the $D_s\bar{D}_s^*$ and $\Lambda_c\bar{\Lambda}_c$ components seem playing minor roles,
respectively.
In addition, the natural values for $a\simeq -1$~fm and $r\simeq -0.5$~fm in the case of the $\psi(4260)$
are compatible with its interpretation in Refs.~\cite{Wang:2013cya,Wang:2013kra} as a  $D_1\bar{D}$ molecular state.  %O-1

\section*{Acknowledgements}
ZHG would like to thank En Wang for the informative communication on
the $\chi_{c1}(4140)$. This work is funded in part by the NSFC under
grants No.~11575052, 11805012, the Natural Science Foundation of
Hebei Province under contract No.~A2015205205, and the MINECO (Spain)
and FEDER(EU) grant FPA2016-77313-P.

\end{document}